\documentstyle[multicol,aps,prb,epsf]{revtex}
\begin{document}

\draft

\title{Polar surface engineering in ultra-thin MgO(111)/Ag(111) 
--- possibility of metal-insulator transition and magnetism}

\author{
Ryotaro Arita$^1$, Yoshiaki Tanida$^2$, Shiro Entani$^3$, 
Manabu Kiguchi$^3$, Koichiro Saiki$^3$, and Hideo Aoki$^1$
}

\address{$^1$Department of Physics, University of Tokyo, Hongo,
Tokyo 113-0033, Japan}
\address{$^2$Fujitsu Laboratories Ltd., Atsugi, Kanagawa 243-0197, Japan}
\address{$^3$Department of Complexity Science \& Engineering,
University of Tokyo, Hongo, Tokyo 113-0033, Japan}

\date{\today}

\maketitle

\begin{abstract}
A recent report  [Kiguchi {\it et al.}, Phys. Rev. B {\bf 68}, 115402 (2003)] 
that the (111) surface of 5 MgO layers 
grown epitaxially on Ag(111) becomes metallic 
to reduce the electric dipole moment raises a question of 
what will happen when we have fewer MgO layers.  Here we have 
revealed, first experimentally with electron energy-loss spectroscopy, 
that MgO(111) remains metallic even when one-layer thick, 
and theoretically with the density functional theory 
that the metallization should depend on the nature of the substrate.  
We further show, with a spin-density functional 
calculation, that a ferromagnetic instability may be expected 
for thicker films.
\end{abstract}

\medskip

\pacs{PACS numbers: 79.60.jv, 73.20.-r, 68.55.-a}

\begin{multicols}{2}
\narrowtext
The problem of polar surfaces, i.e., surfaces of ionic 
crystals where all the outermost atoms are either anions or cations, is 
of fundamental as well as technological interest.   
A textbook description is that 
a polar surface, which would accompany a spontaneous 
appearance of a macroscopic electric field, should be unstable, and 
the surface usually reconstructs itself to 
wash out the electric field.  
While polar surfaces have been envisaged to be 
a unique playground for catalysis, unusual adhesion, 
etc., due to their high reactivity, one distinct 
feature is the possibility of metallization 
of the surface layer that would imply a purely two-dimensional 
metal, on which we focus in the present study. 

The physics behind the metallization of polar surfaces 
is as follows.  A slab having polar surfaces is equivalent 
to a stack of dipole layers, which produce the bulk electric field.  
Since this state has an infinite surface energy, 
a sheet of compensating charges must be formed 
at the top and bottom surfaces.  The system can accomplish this 
in two ways: either the surface is reconstructed to introduce, e.g., 
systematic vacancies, or a charge transfer may occur across the bulk 
and the surface without reconstructions.
In the latter case the sheets of surplus charges imply that each surface 
becomes metallized.

The proposal of the metallized polar surface 
has long been investigated theoretically\cite{Nosker70}.  Specifically, 
the unreconstructed MgO(111) surface, 
which consists of a top O$^{2-}$ surface and a bottom Mg$^{2+}$ surface 
(type-III surface in the classification 
by Tasker\cite{Tasker79}), 
a quantum mechanical calculation has been performed with the discrete 
variational(DV)-X$_\alpha$ 
method\cite{Adachi78} for a finite 
cluster\cite{Tsukada82}, a semi-empirical Hartree-Fock\cite{Pojani97}, 
or a first-principles density functional 
calculation\cite{Goniakowski99,Goniakowski02} for a slab.  
The metallization has been confirmed in these studies.

On the other hand, the problem has a long history of experimental 
investigation as well, but only recently do we have definite results.  
Namely, 
various attempts to grow polar surfaces of rocksalt-structure compounds 
with unreconstructed polar surfaces have proved to be a 
difficult task. It has in fact been shown theoretically that 
reconstructions cannot be avoided unless we introduce 
hydroxylation\cite{Pojani97,Refson95} or adsorption of 
metals\cite{Goniakowski99,Goniakowski02,Benedek96}.  
If we turn to oxide surfaces, they are even more difficult to prepare 
than ionic crystal surfaces, 
because oxygen deficiencies tend to occur and hinder 
spectroscopy for oxide surfaces even when non-polar.  
Now, recent advances in fabrication techniques for ultrathin 
oxides\cite{Henrich94,Noguera96} is enabling us 
to prepare atomically-controlled oxide single crystal films,\cite{Noguera2000} 
but the unreconstructed polar surfaces remained a challenge. 

Recently, however, three of the present authors\cite{Kiguchi2003}
have succeeded in growing an 
MgO(111) 1$\times$1 surface by alternate adsorption
of Mg and O$_2$ on Ag(111), which has enabled them to have a 
unreconstructed polar surface (which turned out to be 
metastable, transforming to more stable ones after annealing).  
The electronic structure of 5 MgO layers (or 
10 monolayers in another nomenclature) of MgO(111), as probed with 
electron energy-loss spectroscopy (EELS) and ultraviolet
photoemission spectroscopy (UPS), shows that the surface is 
indeed unreconstructed with a nonzero density of states at the Fermi energy 
($E_F$) indicative of a metal.  

This raises an interesting question of what should happen 
to the metallization when we have only few monolayers: 
As the film thickness decreases the macroscopic 
electrostatic potential due to the surface charge should decrease, 
so the competition between the metallic state and
the insulating state accompanied by an electric field 
should become more subtle and interesting.  
The present study exactly addresses this problem.

Here we show, first experimentally, that even one layer of MgO(111) 
on Ag(111), as probed with EELS, is metallic.  
We then theoretically show with an {\it ab-initio} calculation that 
the system is indeed metallic.  We further predict that 
the nature of the surface should depend on the 
nature of the substrate, i.e., the surface should be insulating 
for substrate with larger lattice constants.  
We finally show, with a spin-density functional calculation, 
that the surface, whose local density of states is large at the 
surface for thicker films, can exhibit a ferromagnetic instability.

Let us start with the experimental result.  
As for the sample preparation we follow Ref.\cite{Kiguchi2003}.
Namely, the MgO film was grown by alternate adsorptions of Mg and O$_2$ on 
the Ag substrate with temperature of 300K, where 
the whole experiments were performed in a ultrahigh vacuum.  
The formation of a clean 1$\times$1 MgO film on the substrate 
is confirmed from a sharp reflection high energy electron 
diffraction (RHEED) pattern, while 
the Auger-electron spectroscopy detects no contamination.  

\begin{figure}
\begin{center}
\leavevmode\epsfysize=40mm \epsfbox{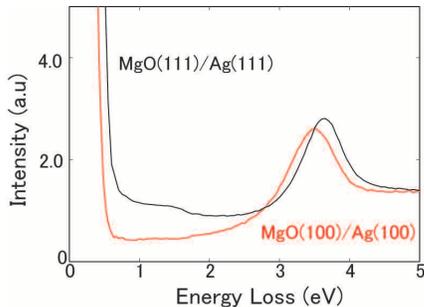}
\caption{EELS 
result, with the primary-electron energy of 60 eV, 
for the one MgO(111) layer on Ag(111) 
as compared with that for MgO(100) on Ag(100).}
\label{exp}
\end{center}
\end{figure}

The electronic structure is then probed {\it in situ} with EELS.
The result is displayed in Fig.\ref{exp} for one 
layer of MgO(111) on Ag(111) as compared with 
MgO(100)\cite{Kiguchi2003} of the same thickness  grown on 
Ag(100).  We can immediately see that the 
MgO(111) has a long and substantial tail for 
the energy loss $\lesssim$ 2.5 eV in a marked contrast with 
the result for MgO(100) 
(while the peak around 4 eV, visible in both cases, originates 
from the surface plasmon of Ag substrate).
The qualitative feature observed here for one layer of MgO on Ag 
is quite similar\cite{commentEL0} 
to those of five layers of MgO on Ag, reported in Ref.\cite{Kiguchi2003}.  
So we conclude that 
one layer of MgO(111) on Ag already has, surprisingly enough, 
a nonzero density of states around $E_F$ suggestive of a metallic 
surface.

We have then performed a first-principles electronic structure 
calculation in the
framework of the generalized gradient approximation (GGA) 
based on the density functional theory.
We adopt the exchange-correlation functional
introduced by Perdew et al.\cite{Perdew1996}
We employ ultra-soft pseudopotentials\cite{Vanderbilt90,Laasonen93} 
for Mg and O, and a norm-conserving soft pseudopotential for Ag, both in 
separable forms. The cut-off energy of the plane-wave 
expansion for the wave function is taken to be 25.0 Ry. 
The atomic configurations and the corresponding electronic 
ground states are obtained with the 
conjugate gradient scheme\cite{Yamauchi1996}.

\begin{figure}
\begin{center}
\leavevmode\epsfysize=40mm \epsfbox{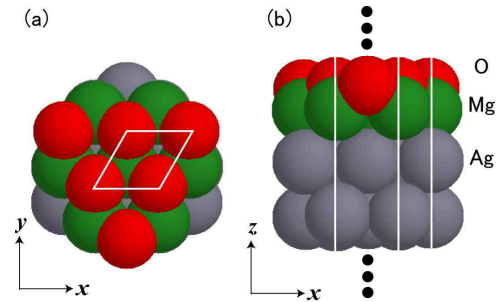}
\caption{
Top(a) and side(b) views of 
the atomic configuration considered in the present study.
The unit cell is indicated by white lines.
}
\label{atom}
\end{center}
\end{figure}

The optimized lattice constants in the bulk 
obtained in the present calculation are 
4.11 (against the experimental 4.09) \AA\ for Ag and  4.22 (4.21) \AA\ for MgO. 
We then introduce a slab model, where we put a (111)-directed 5 
Ag layers sandwiched from top and bottom by MgO layers. 
The structural optimization is done as follows.  
We put Mg atoms on the hollow sites of the outermost 
Ag atoms (see Fig.\ref{atom}), since the total energy 
is found to be lower by $\sim$ 0.1eV 
per unit cell than when we put them on the atop sites.  
The bulk lattice constants of Ag and MgO are slightly 
different (4.09 against 4.21), and 
we assume in the calculation that the lattice constant of the Ag substrate 
does not change at the interface.  So 
we take the same unit-cell size as the bulk of Ag in the $xy$-plane, 
and optimize the positions of the atoms.  
The size of the whole slab in $z$ direction is set to be 
large enough (47.46 \AA).  
Then we put oxygen atoms to have one layer of MgO(111).  
We repeat this for $\geq 2$ layers for multi-layer cases.  
While larger unit-cell sizes in $xy$ would be 
necessary to describe reconstructed surfaces, we take 
1 $\times$ 1 unit cell here since we focus on the unreconstructed case.  
The reasoning is as follows.  The unreconstructed surface is experimentally 
observed as the 1 $\times$ 1 RHEED pattern, but the state 
is metastable.  So if we want to focus on such a state we can perform 
a density functional calculation in the prefixed unit cell. 

\begin{figure}
\begin{center}
\leavevmode\epsfysize=50mm \epsfbox{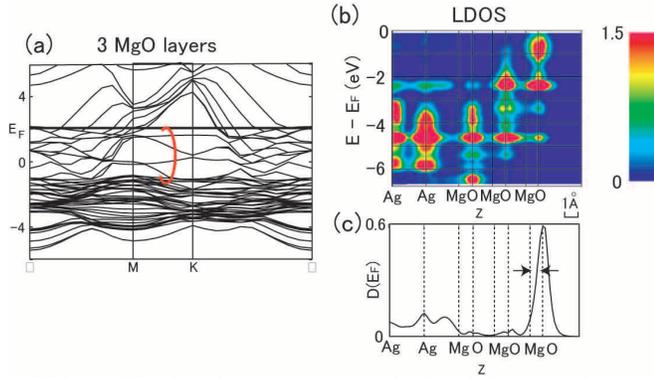}
\caption{The band structure (a), the local density of states (LDOS) integrated 
over both $x$ and $y$ directions displayed as a function of 
energy and $z (\perp$ interface) (b), 
and the LDOS just at $E_F$ as a function of $z$ (c). 
The red mark in (a) indicates MgO-charactered bands, while 
vertical dotted lines in (c) represent the optimized positions 
of atoms, with the separation between the outermost O and Mg 
highlighted by arrows.
}
\label{3layer}
\end{center}
\end{figure}

Let us start with the result for three MgO layers on Ag.  
Figure \ref{3layer} shows the band structure along with the
local density of states(LDOS).  
The LDOS at $E_F$, plotted in Fig.\ref{3layer}(c), 
is calculated by $\sum_{i}|\phi_i(x,y,z)|^2$, 
where the summation is taken over the eigenstates (labelled by $i$) 
having the eigenenergies $E-0.125<E_i<E+0.125$ eV.  
The LDOS does not change significantly when the energy window (0.25 eV here) 
is changed to 0.5 eV.  
The number of sampled k points in Figs.\ref{3layer}(b)(c) 
is 8 with the Monkhorst-Pack
method for the integration over the Brillouin zone\cite{Monkhorst},
where the bands are fitted to sinusoidal forms and the tetrahedron 
method is employed.  The result in fact changes little when the 
number is increased to 18.   

We can see in the energy-resolved LDOS (Fig.\ref{3layer}(b)) that, 
while the LDOS at $E_F$ is 
small, LDOS is large around $E \simeq E_F-5$ eV, 
which originates from 3$d$ levels of Ag.  
If we turn to the LDOS at $E_F$ in Fig.\ref{3layer}(c), 
a notable peak is seen at the outermost oxygen, which suggests 
that the surface is in fact metallic. 
This is a tail (in real space) of the bands originating 
from the MgO (marked in k space in Fig.\ref{3layer}(a)).   
We have obtained a similar result for the case of two MgO layers (not shown).
This is quite consistent with the 
experimental result\cite{Kiguchi2003} that unreconstructed polar surface 
becomes metallic if the MgO film is $\gtrsim 5$ MgO layers, and 
also with theoretical 
studies\cite{Tsukada82,Pojani97,Goniakowski99,Goniakowski02}. 

\begin{figure}
\begin{center}
\leavevmode\epsfysize=50mm \epsfbox{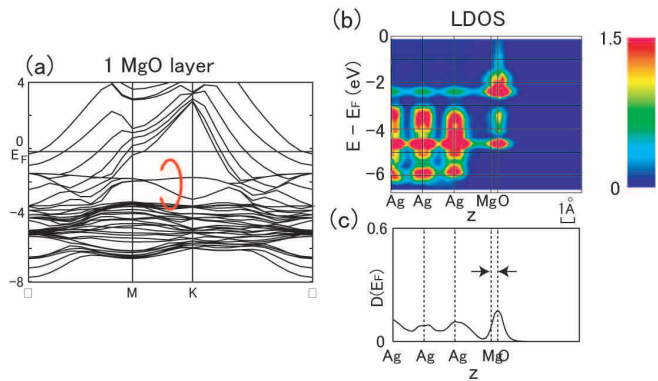}
\caption{A plot similar to Fig.\ref{3layer} for one MgO(111) layer.}
\label{1layer}
\end{center}
\end{figure}

Now, for the thinnest possible (one) MgO layer 
in Fig.\ref{1layer}, we can see in the LDOS at $E_F$ (c) 
that the peak at the O site continues to exist, which 
indicates that the one MgO(111) layer 
is still metallic as consistent with the experiment described above.  
To be more precise, the tail in real space still extends to the 
surface for the bands originating 
from MgO although they [marked in Fig.\ref{1layer}(a)] 
are smaller in number for thinner films.   
The peak at the O site is seen to be 
much smaller in hight than in Fig.\ref{3layer}(c), so that 
the charge redistribution occurs less completely for one MgO layer 
than for three layers.  
Experimentally it is difficult to 
quantify the difference between one- and 
five-layer cases in the present EELS, but other techniques such as 
photoemission spectroscopy or energy loss spectroscopy 
with higher energy resolutions should detect the difference.

So a small electric field may remain for the 
incomplete charge redistribution, but we will have to implement 
a theoretical method to quantify the electric 
field\cite{Goniakowski99,Goniakowski02}.  
Qualitatively, however, we can understand 
the incomplete charge redistribution for one MgO layer 
as follows.
We first note in Fig.\ref{1layer}(c) that the outermost O atoms 
sink deep into the Mg layer.  Namely, the optimized 
separation, $d$, in $z$-direction 
between the O and Mg atoms on the surface is 0.50 \AA, which is 
much shorter than that for the case of three MgO layers (with $d=$ 0.90\AA).
This suggests that the electric field induced by the surface charge is
small for one MgO layer.
In other words, the energy loss due to the lattice distortion 
($\Delta_L$) and that due to the buildup of the electric field
($\Delta_E$) are relatively moderate for one MgO layer, 
so that the energy cost due to
the charge redistribution ($\Delta_C$) may exceed 
$\Delta_L+\Delta_E$.  

This observation leads us to propose here an interesting phenomenon:  
if we can change the value of $d$ which dominates $\Delta_L+\Delta_E$, 
we can change the relative magnitudes of 
$\Delta_L+\Delta_E$ and $\Delta_C$. 
This implies that we should be able to induce a 
{\it metal-insulator transition} for
the polar surface by controlling $d$. 
One of the simplest ways to change the value of $d$ 
is to vary the lattice constant of the substrate. Namely, if we employ 
a substrate having a larger lattice constant, 
the separation between the hollow sites becomes larger, 
which in turn increases the distance between Mg atoms in $xy$-plane.   
This will make O atoms sink more deeply into the Mg layer, and 
a smaller $d$ will favor an insulating surface.  

\begin{figure}
\begin{center}
\leavevmode\epsfysize=50mm \epsfbox{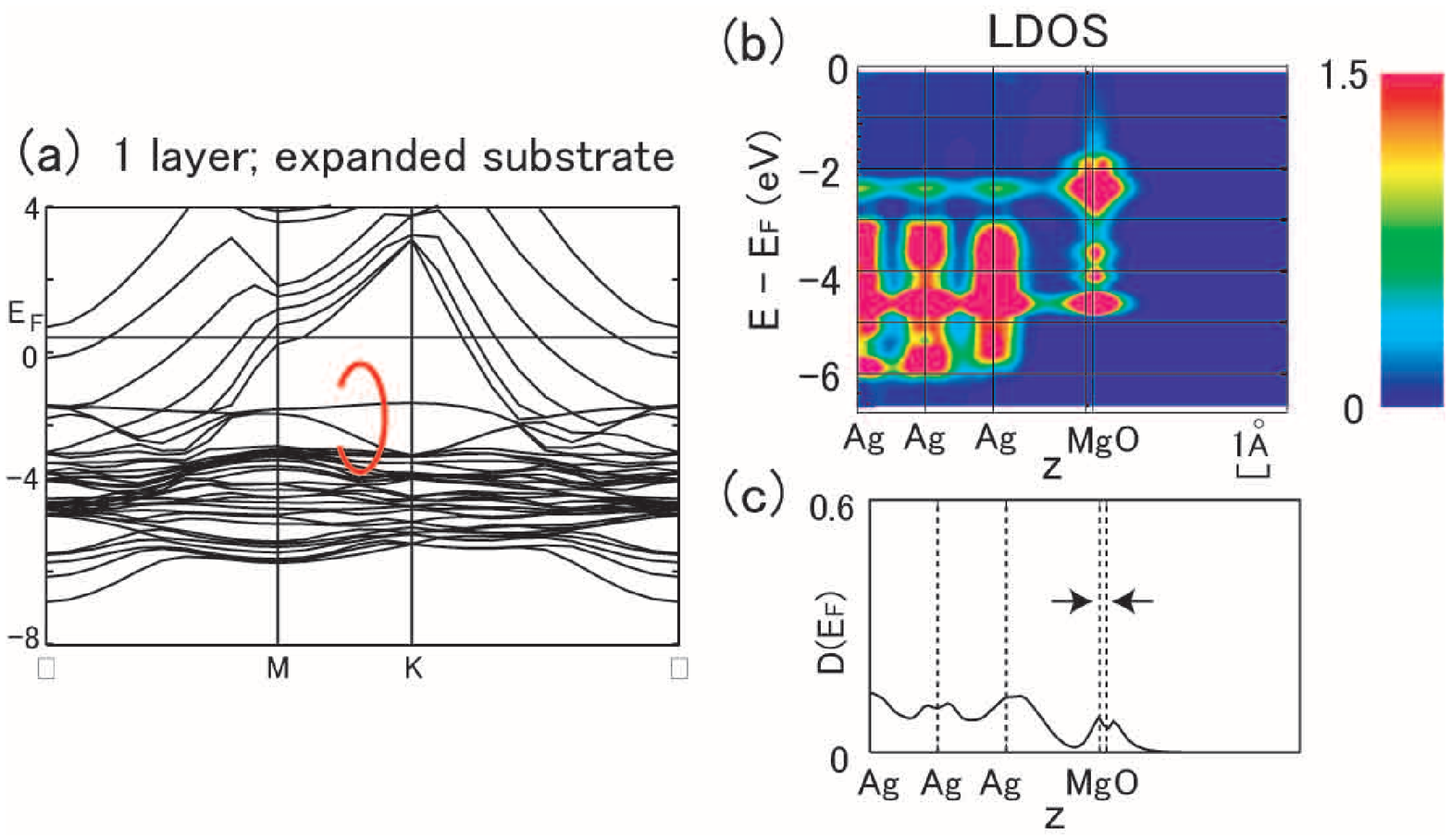}
\caption{A plot similar to Fig.\ref{3layer} for one MgO layer with a 
larger lattice constant (4.23 \AA) of Ag than that in Fig.\ref{1layer}.}
\label{large}
\end{center}
\end{figure}

\begin{figure}
\begin{center}
\leavevmode\epsfysize=50mm \epsfbox{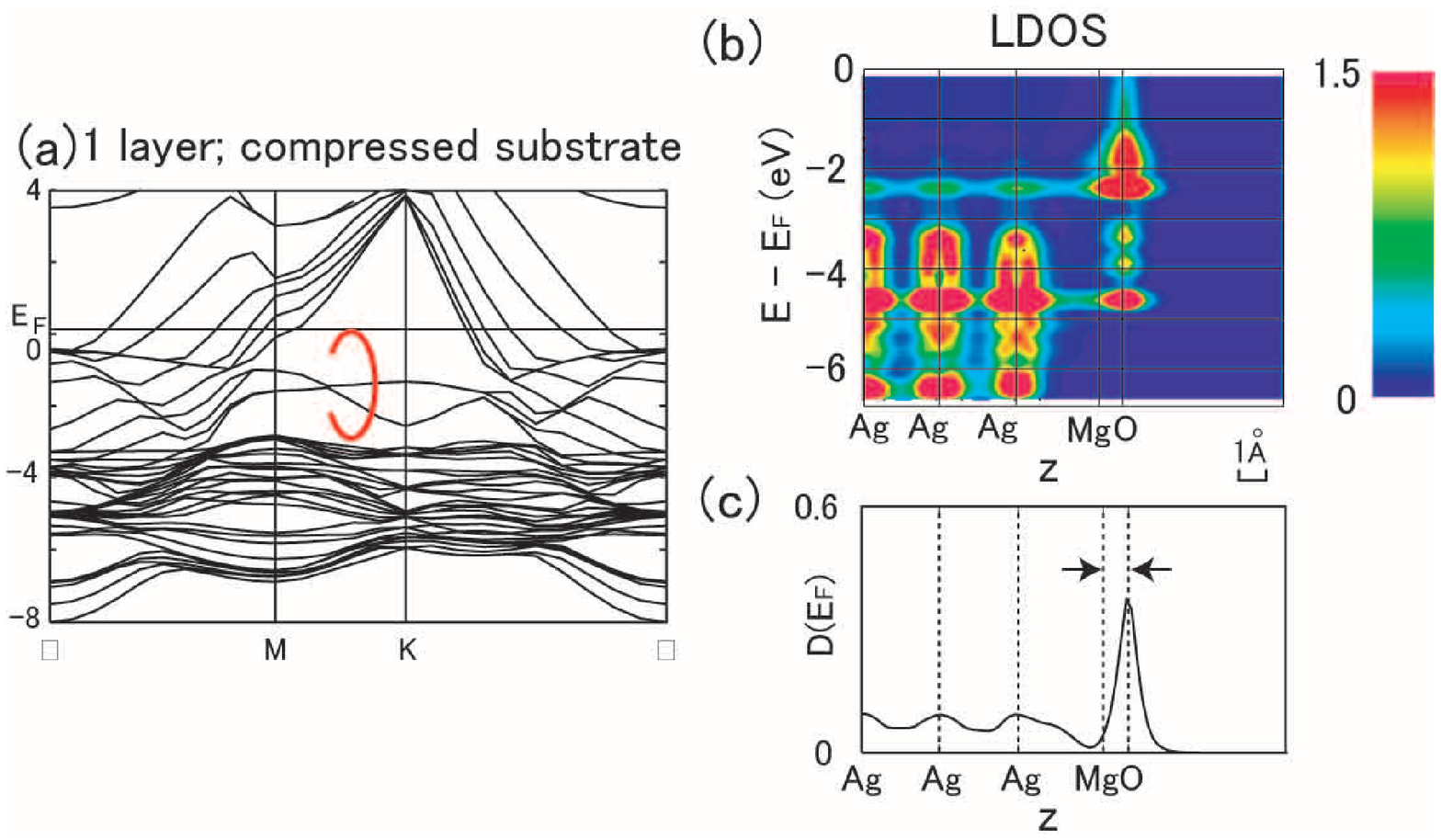}
\caption{A plot similar to Fig.\ref{3layer} for one MgO layer with a 
smaller lattice constant (3.90 \AA) of Ag than that in Fig.\ref{1layer}.}
\label{small}
\end{center}
\end{figure}

To confirm that the MgO surface becomes insulating for 
substrates having large lattice constants, 
we have calculated the band structure
and LDOS for the interface system where we artificially 
make the lattice 
constant larger (4.23\AA\, Fig.\ref{large}) or smaller 
(3.90\AA\, Fig.\ref{small}). 
We can see that the larger lattice constant indeed 
makes the LDOS around $E_F$ at the outermost O (Fig.\ref{large}) 
notably smaller than in Fig.\ref{1layer}, 
while the quantity becomes larger for the compressed substrate 
(Fig.\ref{small}). 
This shows that the system indeed resides
in the vicinity of the metal-insulator transition,
so it should be feasible to control the transition.

 From the behavior of the LDOS at $E_F$ we can look into an even more 
fascinating phenomenon, i.e., 
the possibility of a magnetic instability 
for the polar surface.  As we have shown, the LDOS at $E_F$ 
is large for the outermost oxygen layer in thicker MgO films, 
where O atoms do not sink into the Mg layer.  
When the LDOS is large enough, we may expect a ferromagnetic 
instability.  In fact, Goniakowski{\it et al.}\cite{Goniakowski99}
have already shown, by means of a spin density functional
calculation for a five atomic-layer slab 
(O/Mg/O/Mg/O with unbalanced numbers of cations and anions), 
that the unreconstructed polar surface of MgO
has a ferromagnetic instability. However, the structure of 
MgO (especially the value of $d$) is not optimized in that study, 
so the LDOS at $E_F$ should be overestimated.

\begin{figure}
\begin{center}
\leavevmode\epsfysize=50mm \epsfbox{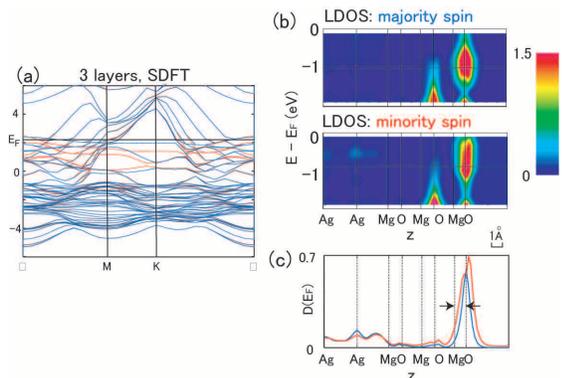}
\caption{A plot similar to Fig.\ref{3layer}, where 
the spin density functional theory with GGA is adopted to 
calculate spin-resolved band structures.  
Blue (red) lines in (a,c) represent the majority (minority) spin.
}
\label{spin}
\end{center}
\end{figure}

So we have performed a spin density functional calculation
for one-layer and three-layer MgO(111) on Ag with the optimization for $d$ 
to study the magnetic instability.  
The result shows that the three-layer system 
does have a ferromagnetic ground state, although 
the magnetization (the difference between the numbers of
opposite spins per unit cell) is 
small ($\sim$ 0.1).  Figure \ref{spin} displays the band
structure and the LDOS, where an exchange splitting 
($\sim 0.5$ eV) is seen for the bands lying $\sim 2$ eV below $E_F$.   
One MgO(111) layer is paramagnetic as expected.  

One thing we have to note is that the unit cell in the 
present work as well as in Ref.\cite{Goniakowski99} is 1$\times$1 in the
$xy$-plane, so that it cannot describe other magnetic states 
such as antiferromagnetic ones.  
Thus a calculation with larger unit cells will be required 
to really confirm whether the ground state can become magnetic, 
which is an important future problem. 

RA and YT would like to thank J. Nakamura for providing them a 
pseudo-potential for Ag. The GGA calculation was performed with TAPP 
(Tokyo ab-initio program package), for which RA received 
technical advices from Y. Suwa.  Numerical calculations
were performed on SR8000 in ISSP, University of Tokyo.
This research was partially supported by a grant-in-aid for creative 
scientific researches from the Japanese Ministry of Education.

\end{multicols}

\begin{references}
\bibitem{Nosker70} R.W. Nosker, P. Mark, and J.D. Levine, Surf. Sci. {\bf 19}, 
291 (1970).
\bibitem{Tasker79}P.W. Tasker, J. Phys. C. {\bf 12}, 4977 (1979). 
\bibitem{Adachi78} H. Adachi, M. Tsukada, and C. Satoko, J. Phys. Soc. Jpn
{\bf 45}, 875 (1978).
\bibitem{Tsukada82}M. Tsukada and T. Hoshino, J. Phys. Soc. Jpn. {\bf 51}, 
2562 (1982).
\bibitem{Pojani97} A. Pojani, F. Finocchi, J. Goniakowski, and C. Noguera,
Surf. Sci. {\bf 387}, 354 (1997).
\bibitem{Goniakowski99} J. Goniakowski and C. Noguera, Phys. Rev. B
{\bf 60}, 16120 (1999). 
\bibitem{Goniakowski02} J. Goniakowski and C. Noguera, Phys. Rev. B
{\bf 66}, 085417 (2002). 
\bibitem{Refson95}K. Refson, R.A. Wogelius, D.G. Fraser, M.C. Payne,
M.H. Lee, and V. Milman, Phys. Rev. B {\bf 52}, 10823 (1995).
\bibitem{Benedek96}R. Benedek, M. Minkoff, and L.H. Yang, 
Phys. Rev. B {\bf 54}, 7697 (1996).
\bibitem{Henrich94}V.E. Henrich and P.A. Cox,
{\it The Surface Science of Metal Oxides} 
(Cambridge Univ. Press, Cambridge, 1994).
\bibitem{Noguera96}C. Noguera, {\it Physics and Chemistry of Oxide
Surfaces} (Cambridge Univ. Press, Cambridge, 1996).
\bibitem{Noguera2000} See, for a review, 
C. Noguera, J. Phys. Cond. Matt. {\bf 12}, R367 (2000).
\bibitem{Kiguchi2003}M. Kiguchi, S. Entani, K. Saiki, T. Goto, and
A. Koma, Phys. Rev. B {\bf 68}, 115402 (2003).
\bibitem{commentEL0} 
The intensity does not vanish for energy loss $\rightarrow$ 0 for MgO(111) 
unlike in \cite{Kiguchi2003}, 
but this is due to a thinner MgO for which 
the Ag substrate contributes to the signal.

\bibitem{Perdew1996}J.P. Perdew, K. Bruke, and Y. Wang, Phys. Rev. B
{\bf 54}, 16533 (1996).
\bibitem{Vanderbilt90}D. Vanderbilt, Phys. Rev. B {\bf 41}, 7892 (1990).
\bibitem{Laasonen93}K. Laasonen, A. Pasquarello, R. Car, C. Lee, 
and D. Vanderbilt,  Phys. Rev. B {\bf 47}, 10142 (1993).
\bibitem{Yamauchi1996}J. Yamauchi, M. Tsukada, S. Watanabe, and O. Sugino,
Phys. Rev. B {\bf 54}, 5586 (1996).
\bibitem{Monkhorst} H.J. Monkhorst and J.D. Pack, Phys. Rev. B
{\bf 13}, 5188 (1976).
\end{references}
\end{document}